\documentclass{article}

\usepackage{PRIMEarxiv}

\usepackage[utf8]{inputenc} 
\usepackage[T1]{fontenc}    
\usepackage{hyperref}       
\usepackage{url}            
\usepackage{booktabs}       
\usepackage{amsfonts}       
\usepackage{nicefrac}       
\usepackage{microtype}      
\usepackage{lipsum}
\usepackage{fancyhdr}       
\usepackage{graphicx}       
\graphicspath{{media/}}     

\title{Perceived personality state estimation in dyadic and small group interaction with deep learning methods
\thanks{\textit{\underline{Citation}}: 
\textbf{Authors. Title. Pages.... DOI:000000/11111.}} 
}

\author{
  Kristian Fenech\\
  Department of Artificial Intelligence \\ 
  Eötvös Loránd University \\
  Budapest, Hungary\\
   \And
   Ádám Fodor \\
  Department of Artificial Intelligence \\ 
  Eötvös Loránd University \\
  Budapest, Hungary\\
     \And
  Sean P. Bergeron \\
  Department of Artificial Intelligence \\ 
  Eötvös Loránd University \\
  Budapest, Hungary\\
     \And
  Rachid R. Saboundji \\
  Department of Artificial Intelligence \\ 
  Eötvös Loránd University \\
  Budapest, Hungary\\
     \And
  Catharine Oertel \\
  EWI, Department of Intelligent Systems \\
  Delft University of Technology \\
  The Netherlands\\
     \And
  András Lőrincz \\
  Department of Artificial Intelligence \\ 
  Eötvös Loránd University \\
  Budapest, Hungary\\
}

\begin{document}
\maketitle

\begin{abstract}
Dyadic and small group collaboration is an evolutionary advantageous behaviour
and the need for such collaboration is a regular occurrence in day to day life.
In this paper we estimate the perceived personality traits of individuals in
dyadic and small groups over thin-slices of interaction on four multimodal
datasets. We find that our transformer based predictive model performs
similarly to human annotators tasked with predicting the perceived big-five
personality traits of participants. Using this model we analyse the estimated
perceived personality traits of individuals performing tasks in small groups
and dyads. Permutation analysis shows that in the case of small groups
undergoing collaborative tasks, the perceived personality of group members
clusters, this is also observed for dyads in a collaborative problem solving
task, but not in dyads under non-collaborative task settings. Additionally, we
find that the group level average perceived personality traits provide a better
predictor of group performance than the group level average self-reported
personality traits.
\end{abstract}

\section{Introduction}

How we express our personality to others is influenced by many underlying
factors such as our relationship to those around us, the social situation in
which we are in and our desired intent for the interaction. When describing the
personality of an individual, it is common to utilise a trait based description
of personality~\cite{mccrae1992introduction}. Within this trait based
description, an individual's personality is composed of the so-called big-five
factors: \textbf{O}penness, \textbf{C}oncientiousness, \textbf{E}xtraversion,
\textbf{A}greeableness and \textbf{N}euroticism (OCEAN). These traits can be
obtained in two ways; from self-report or from other-report. In the field of
affective and personality computing the estimation of self-report personality
is commonly known as personality recognition. Whereas the estimation of
other-reported personality is known as personality
perception~\cite{Vinciarelli2014personality}. Personality perception can be
understood as the act of estimating the expression of a persons personality in
the big five traits as viewed by an external observer. The perception of
personality is influenced by a wide set of sources~\cite{ekman1980relative},
specfic attributes of speech such as speech rate~\cite{smith1975effects},
voice quality and intonation~\cite{mohammadi2012speech} have been shown to
influence such estimations. Non-verbal cues also play a significant role in the
perception of personality~\cite{vinciarelli2012nonverbal}. This has been
shown in studies of gaze~\cite{ijuin2020exploring}, back
channels~\cite{blomsma2022backchannel} and body language
cues~\cite{romeo2021predicting}. 

Growing bodies of evidence have continue to support the relationship between
personality and performance across a variety of tasks and both at the level of
the individual~\cite{ziegler2010predicting,poropat2014other} and at the group
level~\cite{prewett2018effects,neuman1999relationship}. While many of these
studies focus on the relationship between self-reported personality traits,
there are many examples in the current literature which outline the importance
of evaluating other-reported perceived personality. While personality traits
tend exhibit long-term stability, they are not
fixed~\cite{allport1955becoming, cobb2012stability,ludtke2009goal}. Studies
have shown that within-person variation of personality traits can show
significant short-term change~\cite{fleeson2001toward}. However, the
resolution at which this variation can be examined is limited by the frequency
of questioning which may be used in current experience sampling
methods~\cite{conner2009experience}. 

With the introduction of the ChaLearn dataset~\cite{lopez2016chalearn}, there
has been significant progress in the area of automated personality perception
from multi-modal sources. We leverage these developments in order to estimate
the perceived personality state of individuals from a multiple audio-video
sources which focus on dyadic and small group interaction. The automated nature
of such methods provide a way to augment the resource and time intensive nature
of experience sampling. 

Through the evaluation of the time varying first-impression like personality,
i.e. estimations of personality in which the perception is not influenced by
earlier predictions which exist outside the thin slice of input under
consideration. We examine the perceived personality state of individuals taking
part in dyadic and group interactions. We examine multiple datasets which focus
on dyadic and small group interaction. In order to analyse the generalisability
of the developed method we make use of the provided perceived personality
annotations present in the MULTISIMO dataset. For the dyadic case we analyse
the UDIVA dataset~\cite{palmero2021context} and for small groups the Emergent
Leader (ELEA)~\cite{sanchez2015elea} and AMI~\cite{mccowan2005ami} meeting corpus.
Both the UDIVA and AMI meeting corpus provide the opportunity to explore the
potential changes in perceived personality over different tasks. Despite the
limitation that the ELEA dataset provides only a single interaction session for
each group, it provides a useful metric of group performance for the completed
task. For all of these datasets we determine a time-averaged set of personality
traits for each individual and evaluate the existence of the emergence of group
personality, that is the convergence of individual personality states at a
group level, as well as changes in expression of the participants personality
across tasks.

In our analysis we first demonstrate the out-of-distribution performance of the
trait estimation model performs similarly to human raters. Using estimates on
perceived personality applied to both small groups and dyads involved in a task
oriented interaction we examine the time-averaged personality states of group
members for the existence of group level clustering and the subsequent
evolution of the time-averaged personality traits across different tasks.
Finally, we compare the predictive capacity of the time-averaged perceived
personality and self-reported personality at the group level on a measure of
group performance in a collaborative problem solving task.

\section{Materials and Methods}

\subsection{Automatic perceived personality estimation}
We estimate the big-five personality traits in a multi-modal
fashion, in which we combine information from audio, textual and visual inputs.
For the audio modality we compute the eGeMAPS features~\cite{eyben2015geneva}
using the OpenSmile~\cite{Eyben2010Smile})software. The textual modality is
obtained from speech transcripts, from which BERT~\cite{devlin2018bert}
embeddings are generated. The final visual modality comprises of facial action
units~\cite{rosenberg2020face} which are obtained using the
OpenFace~\cite{tadas2018openface} software. Action units are used over the
raw RGB video content to avoid influence due to visual factors such as
clothing, location and background objects visible in the video.

\begin{figure}
\centering
\includegraphics[width=1.0\textwidth]{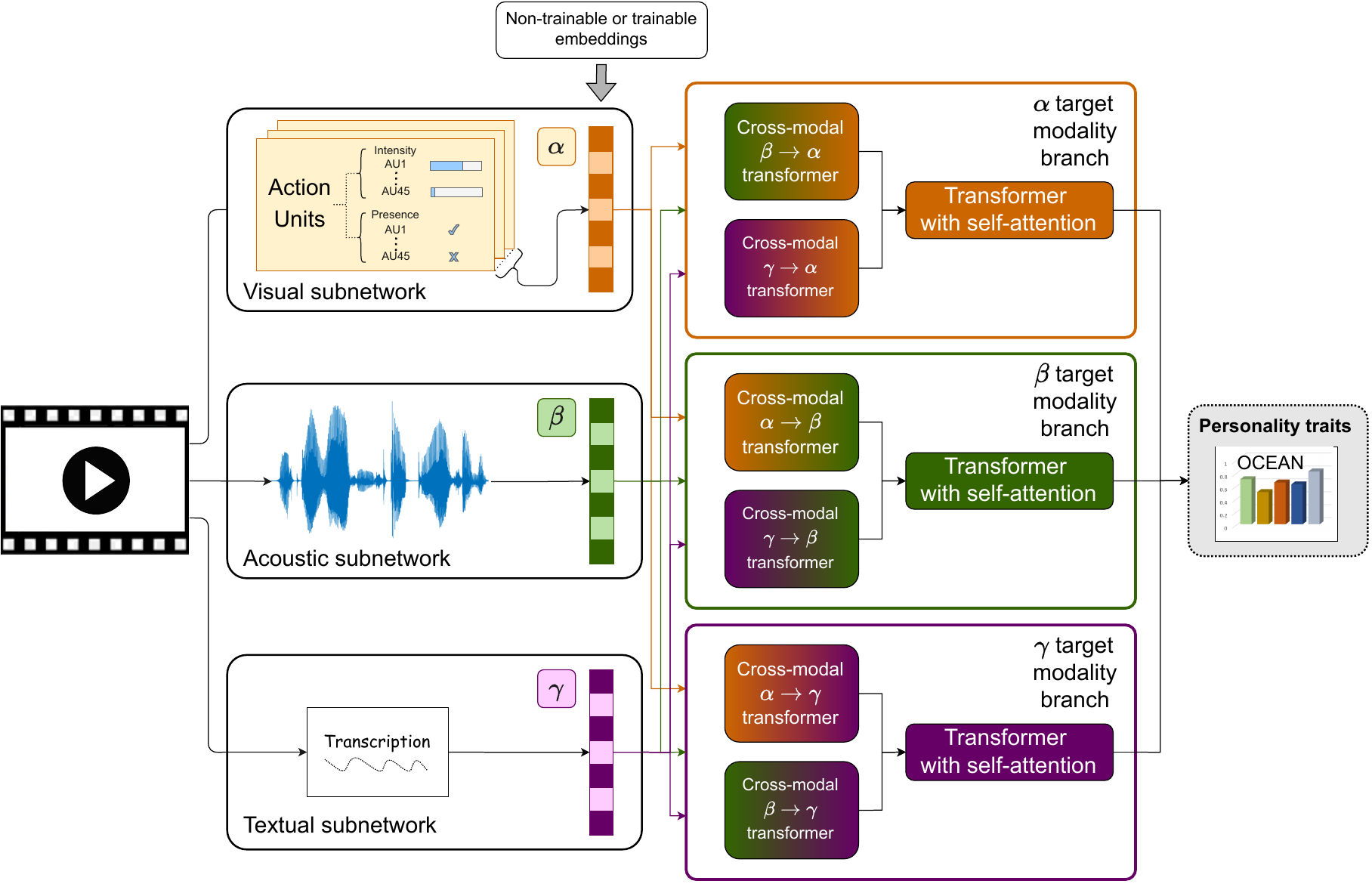}
\caption{Linear Multimodal Transformer architecture: Input embeddings indicated
by darker striped columns can be features derived from the raw data or the
outputs of a pre-trained deep model. Cross-modal transformers translate one
information source (e.g., $\beta$ and $\gamma$, here associated with the
acoustic and textual modality, respectively) to another one (e.g., $\alpha$,
here associated with the visual modality) by learning keys and values of
modality $\beta$ (or $\gamma$) and the queries from the $\alpha$ modality. All
combination of these modalities are utilized by an individual cross-modal
transformer. Such two units with the same target modality are combined by
another transformer network with self-attention to fuse the information pieces
before outputting the predicted scores. Linear attention mechanism is used
within the networks to increase efficiency by reducing computational resources,
i.e., time and memory.}
\label{fig:architecture}
\end{figure}

These features are used with a transformer based model~\cite{vaswani2017attention,
phuong2022formal}. An overview of the architecture is given in
Fig.~\ref{fig:architecture} and full details of the architecture can be found
in~\cite{fodor2022multimodal}. The architecture is built on multi-head
attention units that transform one modality to another. As we utilise three
unique modalities, the model comprises of six cross-modal transformers.

This architecture follows that of~\cite{tsai2019multimodal},
however we replace the quadratic attention modules with linear
versions~\cite{katharopoulos2020transformers}, providing similar performance ~\cite{fodor2022multimodal} while being easier to train. The final output of the
network is the prediction of the big-five personality traits, however we use
emotional stability that refers to the ability that the person can remain
balanced. The low end of our scale corresponds to high neuroticism, i.e., the
person experiences negative emotions often.

\begin{figure}
\centering
\includegraphics{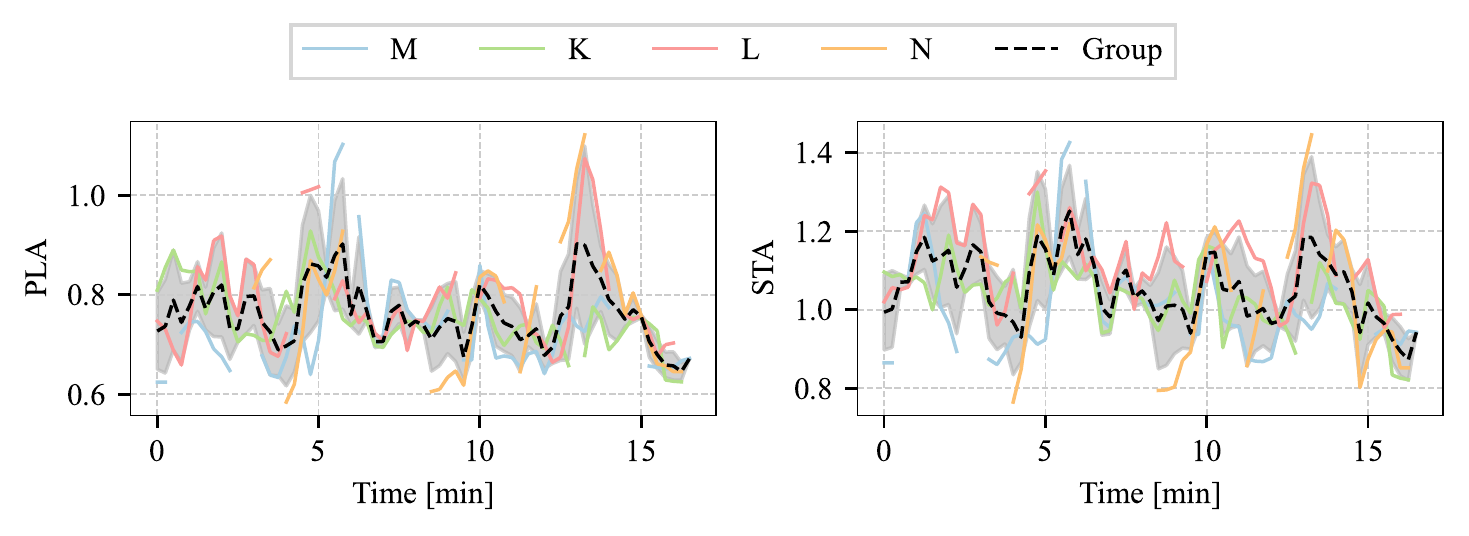}
\caption{Illustrative sample of the temporal changes in the perceived
         personality meta-traits of plasticity (left) and stability (right).
         Coloured curves: Trait estimation for individual participants, where
         each participant is assigned a letter in place of a name for
         anonymisation in the dataset. Missing line segments: non-speaking
         intervals. Dashed line: average over participants. The grey shaded
         area gives the standard deviation of the individual values. Perceived
         personality tend to move together over time.
         }
\label{fig:trajectory}
\end{figure}

Estimation of the big-five traits is made over a sliding window of duration 15
seconds and a stride duration of 1 second. The estimations are then averaged
over a period of 30 seconds, providing a snap-shot of the perceived personality
traits. From the big-five traits we determine a higher-order two-factor
meta-trait representation of personality based on the suggested combininations
of big-five traits as described in~\cite{Digman1997Higher, deyoung2006higher,
deyoung2015cybernetic, strus2014circumplex}. These meta-traits are plasticity
(PLA) and stability (STA), the former combining the traits of openness and
extroversion and the latter conscientiousness, agreeableness and emotional stability.
We express our traits of plasticity and stability as linear combinations of the
relevant big-five traits with equal weights set to unity. An example of the
extracted meta-trait trajectories for one group is shown in
Fig.~\ref{fig:trajectory}.

\subsection{Data sets}
Four data sets were utilised in the study, these were the Emergent Leader
(ELEA), AMI Meeting corpus, UDIVA and MULTISIMO. The ELEA dataset~\cite{sanchez2015elea}
contains video of small groups (3--4 participants) completing a collaborative
task to produce a single ranking of items in terms of importance in a winter
plane crash survival scenario. From this dataset we analysed 17 videos
containing all three (audio, video, text) modalities. This corresponded to
sessions: 12, 14, 16, 17, 21--26, 28, 32, 34, and 36. 

The AMI meeting corpus~\cite{mccowan2005ami} is a multi-modal dataset which
aims to explore small group interactions in meetings. In
these scenario meetings each group takes part in a series of four
meetings with the goal of designing a new remote control. The meeting series starts with an initial kick-off
meeting, followed by meetings on functional design, conceptual
design and finishing with a detailed design meeting, these meetings are labeled
as 'A', 'B', 'C' and 'D', respectively.

Understanding Dyadic Interactions from Video and Audio Signals (UDIVA)
~\cite{palmero2021context} is a multimodal, multiview, non-acted dataset that
consists of face-to-face dyadic interactions. The participants sit at a table,
and they are individually recorded during the sessions. There are 4 different
tasks (Ghost, Talk, Animal, Lego) which vary in the level of collaboration and
competition. This is a multi-lingual dataset, from which we focus only on the 
English language component.

MULTISIMO~\cite{koutsombogera2017multisimo} is another multi-modal dataset
which is triadic, however the majority of interactions are dyadic. In
the dataset, pairs work together to solve a quiz given by the
third participant. We do not include this dataset into the analysis, however,
as this dataset provides annotation of the big-five traits of the perceived
personality estimated by external observers.

\subsection{Data analysis methods}
Evaluation of the trait estimation model, when applied to out-of-distribution
samples (i.e. samples collected at a different time and under different conditions
and environments) is conducted on the MULTISIMO dataset. We determine the inter-rater
correlation factors by evaluating the average of all raters in which we
consider the raters to be randomly selected. A two-one-sided t-test was applied
to the distributions of the absolute error from the mean rating between each
rater and the trait estimation model to determine equivalence of the model
errors and human errors. We conduct this analysis trait-wise, and utilise a
bound determined as the mean absolute error of the human rater for the target
trait. 

In order to evaluate any significance in group clustering we performed a
PERMANOVA test with a pseudo f-ratio test statistic as described
in~\cite{Anderson2001permutation}. The test statistic is determined as $F =
\frac{SS_A / (a-1)}{SS_W / (N - a)}$ where $a$ is the total number of groups in
the study, $N$ is the total number of observations. $SS_{A}$ denotes the
between-group variance. $SS_{W}$ is the within group variance and is calculated
as the sum over all groups of the sum of the squared distance of individuals in
the group to their respective group mean. 

We computed the Maulchy's W for each dependent variable which indicated that
the variances were not equal for all dependent variables in the case of
individuals and for the stability trait for groups. For these dependent
variables we apply the Greenhouse-Geisser correction in our one-way repeated
measure analysis of variance (ANOVA). For our post-hoc analysis we perform
Welch's t-test pairwise between meeting types for the case of AMI and between
task types for UDIVA. For all pairwise t-test's we apply a Holm-Bonnferoni
correction.

For the case of UDIVA, 10 individual participants repeated the tasks with a new
partner. An additional repeated measure ANOVA was conducted on the same
dependent variables, taking the task as the within factor.

Group performance prediction is conducted as a regression task using a Gradient
Boosted Tree method. The performance metric under evaluation is the differences
in ordering of the item importance determined by the group compared to an
expert. Due to the limited amount of data we apply a leave-one-out
cross-validation methodology, and report the average mean-squared error across
all splits.

\section{Results}
Interrater correlation factor ICC(2,k) for the eight external
observers surveyed in the MULTISIMO dataset were determined traitwise.
Using the interpretation of the ICC values as described
in~\cite{koo2016guideline} we observe moderate agreement for the traits of
openness ($0.55$), conscientiousness ($0.62$), and emotional stability ($0.57$). For
extraversion ($0.81$) we find good agreement and we observe poor agreement the
trait of agreeableness ($0.43$), for the meta-traits we see good to moderate 
agreement with plasticity and stability ICC values of $0.80$ and $0.62$ 
respectively.

Comparing the human raters to the predictive model we employed a two one-sided
test (TOST) analysis, for both big-five and meta-traits. This analysis showed 
that for the OCEAN traits, the absolute
error of the model from the human rater average is equivalent to
at least half of the human raters. In the case of
the meta traits, the absolute error of the plasticity is only equivalent with
three of the eight raters and stability with seven out of eight raters.

\begin{figure}
\centering
\includegraphics{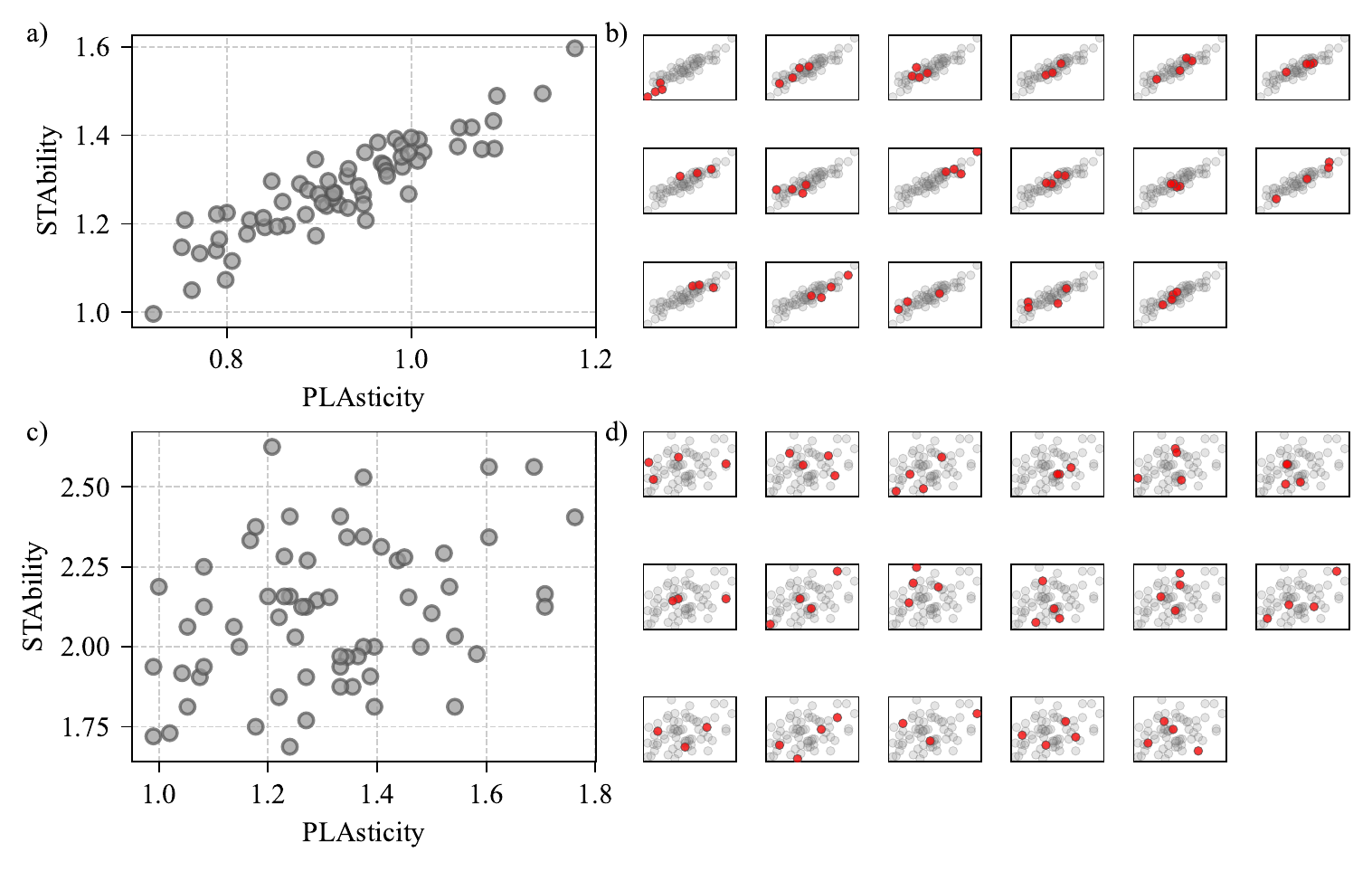}
\caption{Stability and plasticity values. a) first-impression
    personality state averaged over the full session duration. b) Each subplot
    shows the original averages as seen in a). Traits corresponding to the
    members of other group are highlighted by the red circles. c) plasticity
    and stability determined from self reported big-five traits, re-scaled
    between 0 and 1. d) Each subplot shows the original averages as seen in c).
    Traits corresponding to the members of other group are highlighted by the
    red circles. The observed group effect is strong in the first-impression
    personality state.}
\label{fig:pla-sta-dist}
\end{figure}

\begin{figure}[h]
    \centering
    \includegraphics[width=0.8\textwidth]{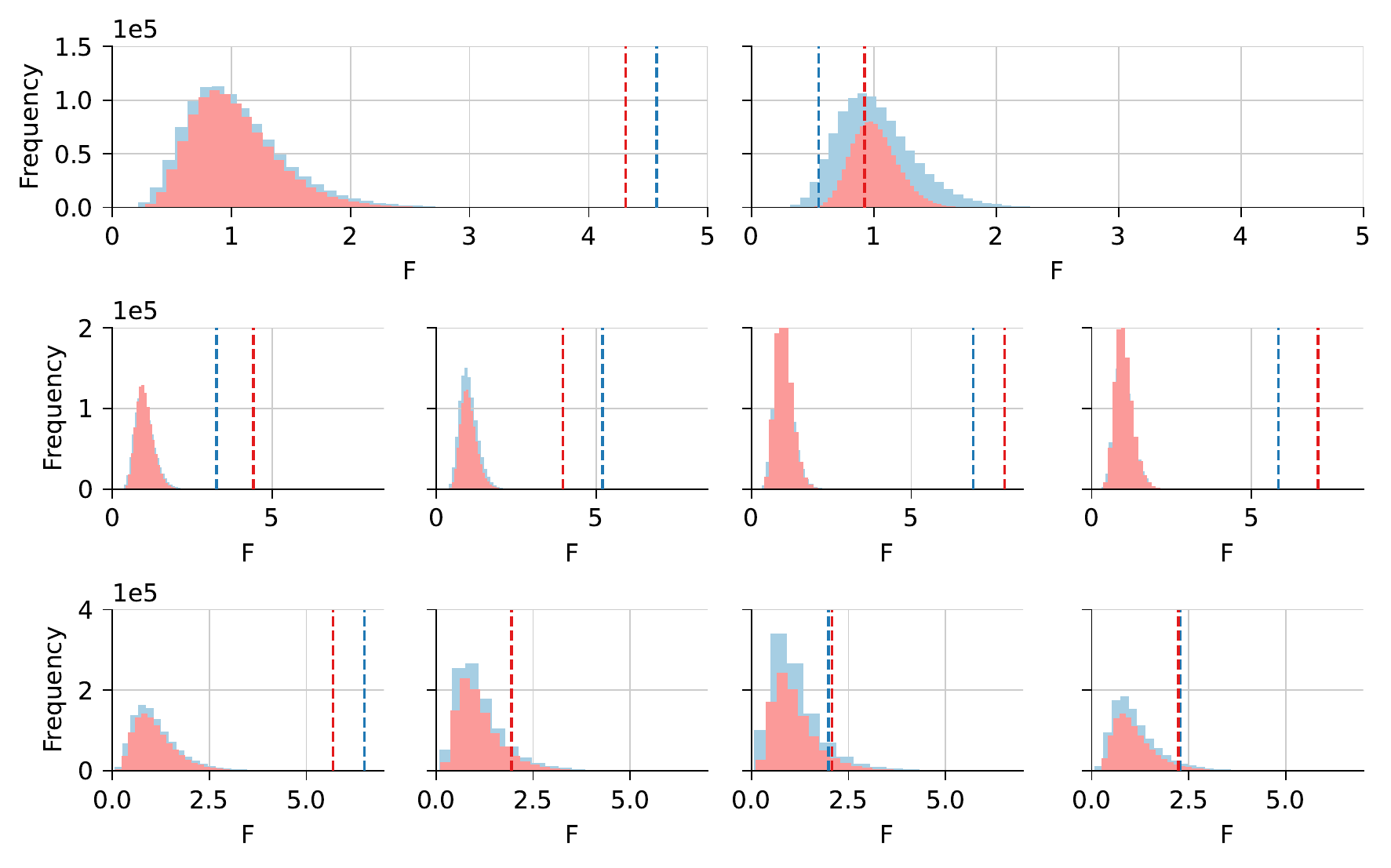}
    \caption{Histograms of the permutation test. For each histogram we show the result for 
    the evaluation of the meta traits in blue and the same evaluation on the raw big five traits
    in red. The top row shows the result for
    the ELEA perceived personality (left) and self-reported (right). The middle row
    shows the results for the four AMI phases from left to right (kick-off, conceptual design,
    functional design, detailed design). The bottom row shows the results for
    UDIVA from left to right (Animals, Lego, Talk, Ghost).}
    \label{fig:img-fig_perm_hist-pdf}
\end{figure}

The distribution of participant session averaged meta-traits are shown for ELEA in
Fig.~\ref{fig:pla-sta-dist}.
For the ELEA and AMI datasets we find the formation of groups to be statistically
significant. We perform the PERMANOVA for all datasets on both the
big-five traits and meta-traits. The resulting histograms
for the PERMANOVA is shown in Fig.~\ref{fig:img-fig_perm_hist-pdf},
with the big-five results in red and the meta-traits in blue. For the ELEA
dataset we observe a test-statistic for the five and two factor personality
models of $F = 4.31$ and $F = 4.57$ respectively, both with $p < .001$.
Similarly for AMI we observe the test-statistics $F_a = 4.41$, $F_b =
3.97$, $F_c = 7.92$, and $F_d = 7.08$ for the big-five and $F_a = 3.26$, $F_b =
5.20$, $F_c = 6.94$, and $F_d = 5.84$ for the meta-traits. For both cases
all meetings have $p < .001$. The result on dyads in UDIVA, shows
only a significant result for the ``animals'' task ($F=6.49, p < .001$).

Repeated measures
ANOVA conducted on the AMI dataset  for individual participants showed a significant difference for all
big-five traits, openness ($F = 16.76$, $p <.001$), conscientiousness ($F =
20.46$, $p <.001$), agreeableness ($F = 5.12$, $p =.002$), extraversion ($F =
19.35$, $p <.001$) and emotional stability ($F = 21.45$, $p <.001$). For the
meta-traits a significant result was found for stability ($F = 3.88$,
$p = .017$). For the plasticity trait a marginally significant result was
obtained ($F = 2.73$, $p=.053$).

Post-hoc pairwise dependent t-tests across meetings showed significant
differences over all combinations of meeting pairs for the trait of openness.
The traits of conscientiousness, agreeableness and emotional stability showed a
non-significant result between meeting B and C. Finally the trait of
extraversion only showed significant difference between sessions A and C and A
and D. No significant differences for the variable of plasticity was observed.
The dependent variable stability showed a significant difference between
meetings A and C.

For the group averaged traits, repeated measure ANOVA,
showed significant differences for the big-five traits of openness ($F =
10.19$, $p < .001$), conscientiousness ($F = 14.64$, $p < .001$), agreeableness
($F= 10.93$, $p < .001$), emotional stability ($F = 12.76$, $p < .001$) and
extraversion ($F=3.15$, $p = .03$). A marginally significant difference was
found for the trait of stability ($F=2.45$, $p=.091$) across meeting types. The
remaining trait of plasticity was not significant.

Post-hoc tests indicated 
the traits of openness and agreeableness
had significant differences for meetings A and B, A and C, and C and D.
Conscientiousness showed significant difference between meeting A and all other
meetings. Extraversion like in the case of individuals showed significant
differences in the fewest number of pairs. For groups this was only between
meeting A and D. All dependent variables based on the meta-traits
showed non-significant results.

Repeated measure ANOVA on individuals in UDIVA returned significant results
for all dependent variables. For the variables related to the meta-traits 
both plasticity and stability were significant with $p < .001$. All
of the big-five traits were also significant with $p < .001$.

For the group averaged traits in UDIVA the repeated measure ANOVA showed only
significant differences for the all dependent variables with $p < .001$.
Post-hoc pairwise t-test 
showed significant differences in  plasticity $p < .001$ for the task pairs of Animals-Lego,
Ghost-Lego and Lego-Talk. For the variable of stability significant differences were
observed with $p < .001$ for all task pairs excluding Animals-Ghost.

In the case of UDIVA, there were 10 participants who repeated the series of
tasks with another partner. A repeated measure ANOVA was performed to examine
the potential effects due to the presence of a new partner. In this case no
significant results for any of the dependent variables on any of the four tasks
was observed.

Having observed across data-sets the emergence of clusters in the perceived
personality we evaluated the predictive capability of both the group-level
perceived meta-traits and the original big-five traits on the metric of group
performance on the ELEA dataset compared to the provided self-reported
personality traits. We obtain a mean-squared error (MSE) of $16.17 \pm 0.45$
with the group average perceived meta-traits and $22 \pm 1.26$ for the big-five
traits. This is compared to a MSE of $43.88 \pm 1.17$ in the case of
self-reported group average meta-traits and $40.68 \pm 1.10$ for the big-five
traits.

\section{Discussion}
By evaluating the equivalence of the
absolute error of the estimated perceived personality to that of the human
rater, where we calculate the error between the estimate and the average rating
of all human raters, we find that in terms
of equivalence plasticity has the fewest equivalences with human raters. For
this trait we can reject the null hypothesis against only 3 out of 8 human
raters. For all other traits we find that there is equivalence within the
prescribed bounds with at least 4 out of 8 humane raters. For openness,
agreeableness and stability we find equivalence with 7 out of 8 human raters, 5
out of 8 for extraversion and neuroticism and 4 out of 8 for conscientiousness.
These results suggest that we can have confidence in the application of the
model on other distinct but similar datasets which we can access video, audio
and text modalities. 

Our results on the remaining datasets are framed by
the following ideas. Firstly, that personality expression is influenced by the
situation experienced by the
individual~\cite{baumert2017integrating,gori2022stable,mcadams2006new,sosnowska2019dynamic,tett2003personality}.
Secondly, in social interactions, participants may mirror their
partners~\cite{chartrand1999chameleon}. We found in the case of small groups
in the ELEA dataset who are solving an object ranking problem and in all AMI
meeting types, the perceived personalities traits of group members formed
clusters. However, the analysis of of the dyadic case present in the UDIVA
data, only shows this effect in ``animals'' task. Of the four tasks undertaken
by the participants of the UDIVA study, only ``animals'' and ``lego'' may be
considered somewhat collaborative.

In the case of ``animals'' the collaboration is achieved by the process of
question and answers made by the participants in order to discover the
participants assigned animal. While the ``lego'' task is collaborative, it does
not require discussion or high levels of engagement between the participants.
In this case the target object to construct is given, and only the process of
construction is required. The ``ghost'' task is purely competitive, with both
participants aiming to optimise their score against the other and ``talk''
present open conversation with no specific goal to be achieved as a group. The
nature of these tasks and the findings suggest that the influence of others in
a group may be felt even in the case of dyads, but the task and context of the
interaction also play a role. The clustering of the perceived personality only
occurring in the predominately collaborative tasks may be seen in the context
of avoidance of social exclusion, which has been shown to effect personality
expression~\cite{dewall2011belongingness}.

In the further analysis of the big-five and meta-traits across meeting types in
the AMI meeting corpus demonstrates that for many of the big-five traits there
is a statistically significant change depending on the meeting type which is
being carried out and has an effect both at the level of the individuals and at
the group level. At both the individual and group level, the perceived
extraversion appeared to be the most stable trait. Conversely, the higher-order
traits as a linear combination of the big-five traits appear to exhibit less
change across meeting types. The stability of the meta-traits does not extend
in our analysis to the case of dyads in UDIVA. This difference in the stability
of the higher-order traits may be due to the fact that in AMI all group tasks
were collaborative and the context (design of a new product) was consistent
between all meetings, with variation in task only being at the level of detail.
Whereas UDIVA featured a wider variety of task goals. Further studies of dyads
with consistent task goals similar to those in AMI may help explain these
variations.

The additional analysis on the UDIVA participants whom repeated the four tasks
with a new partner, showed no significant difference in their perceived
personality between their first and second attempts at each task. However, due
to the low number of participants who had repeated the set of tasks, we could
not draw strong conclusions regarding this outcome.

While we have analysed the time-averaged traits in these studies, there remain
many open questions about the factors which influence the observed dynamics of
the perceived personality and the relationship they have to psychological
processes~\cite{Fleeson2004Moving}. The work of~\cite{saucier2007modifies}
describes a taxonomy of situations, which describe locations, associations,
activities and experienced processes which are relevant to the expression of
personality. Such a taxonomy may be useful in future experiment designs and the
study of these factors may enable prediction of future changes in a individuals
perceived personality state.

In addition to identifying causal factors, there is the question of what group
and within group attributes may be predicted from the perceived personality of
the group. In our study we compare the predictive capability of the perceived
and self-reported traits from the ELEA dataset. We find that the perceived
traits result in a more accurate prediction of group performance. While a large
body of literature exists examining the relationship between self-reported
personality and performance across a variety of
tasks~\cite{halfhill2005group,kramer2014personality,
neuman1999relationship,prewett2018effects}, these results add further support
to the idea that self-report alone may not be sufficient and in fact,
other-rated personality traits may be stronger
predictors~\cite{mount1994validity, smith1967usefulness,poropat2014other}).

\section{Conclusion}
In summary, this paper investigates the perceived personality of individuals in
dyads and small groups estimated using deep-learning based methods, across
multiple datasets. We first determined the perceived personalty traits
represented as the big-five traits, using a multi-modal linear transformer
model evaluated on thin slices of the recorded meetings. We found that the
model used to determine the perceived personality traits performs similarly to
human raters present in the MULTISIMO dataset.

Using these estimates of perceived personality we evaluate the significance of
the group membership with a non-parametric permutation test. For the ELEA and AMI
datasets, we found the clusters centroids to be significantly different between
groups with with $p < .001$, and for UDIVA a significant result was only
observed for a single task out of the four evaluated tasks with $p < .001$. The
task dependent nature of these results suggest that in small groups and dyads
in which the task is collaborative, how personality is expressed clusters
around the group. For the ELEA dataset it was observed that this result
did not occur for the self-reported personality. Additionally we studied how
both group and individual perceived personality traits varied over multiple
tasks, we found that in the case of tasks in which the task does not deviate
significantly, that the group average perceived personality traits demonstrated
fewer significant changes between tasks compared to the individual group
members. However, when the task context is significantly different between
tasks we observe similar amounts of variation in the individuals and in the
group averages.

Finally we evaluated the predictive nature of the perceived traits compared to
self-reported traits for the task of group performance prediction. In the
winter-survival task presented in the ELEA dataset the group average perceived
personality meta-traits of plasticity and stability were a stronger predictor
of the overall group performance than the group average self-reported
meta-traits.


\section*{Author Contributions} 
\textbf{KF} Conceptualization, Formal Analysis, Software, Writing - original
draft. \textbf{AF} Data Curation, Formal Analysis,
Software, Writing - original draft. \textbf{SB} Investigation. \textbf{RS} Data curation, Software.
\textbf{CO} Writing - review \& editing . \textbf{AL} Conceptualization,
Methodology, Writing - original draft, Supervision, Project administration.

\section*{Funding} 
Support by the the European Union project RRF-2.3.1-21-2022-00004 within the
framework of the Artificial Intelligence National Laboratory. This work was
partially supported by the European Commission funded project ``Humane AI:
Toward AI Systems That Augment and Empower Humans by Understanding Us, our
Society and the World Around Us'' (grant \# 820437). The support is gratefully
acknowledged.

\section*{Acknowledgments} 
We would like to thank Sergio Escalera and David Gallardo-Pujol for their helpful 
discussions. The authors also thank Robert Bosch, Ltd. Budapest, Hungary for their
generous support to the Department of Artificial Intelligence.
\bibliographystyle{unsrt}  
\bibliography{references}

\end{document}